Temperature dependence of spin diffusion length in silicon by Hanle-type spin precession

T. Sasaki[1,a], T. Oikawa[1], T. Suzuki[2], M. Shiraishi[3], Y. Suzuki[3], and K. Noguchi

[1]SQ Research Center, TDK Corporation, 385-8555 Nagano, Japan

[2]AIT, Akita Research Institute of Advanced Technology, 010-1623 Akita, Japan

[3]Graduate School of Engineering Science, Osaka University, 560-8531 Osaka, Japan

Abstract

   The Hanle-type spin precession method was carried out in association with non-local (NL) magnetoresistance (MR) measurement using a highly doped ($5 \times 10^{19}$ cm$^{-3}$) silicon (Si) channel. The spin diffusion length obtained by the Hanle-method is in good agreement with that of the gap dependence of NL signals. We have evaluated the interface and bulk channel effects separately, and it was demonstrated that the major factor of temperature dependence of NL signals originates from the spin polarization reduction at interface between the tunnel barrier and silicon.

[a]E-mail address: tomosasa@jp.tdk.com

Semiconductor spintronics has a large potential for electronic (and spin) devices. From the viewpoint of the application silicon (Si) can be regarded as an ideal material, because it has a small spin orbit interaction as well as crystal inversion symmetry; consequently a long electron spin diffusion length and spin lifetime are anticipated.[1] By using the long spin diffusion length predicted in Si, logic devices such as Spin MOSFET have been proposed.[2] Recently, Appelbaum *et al.* have demonstrated the coherent spin transport through a 2 mm distance in a Si wafer by using hot-electrons.[3, 4, 5] According to Ref.4, the spin lifetime in non-doped Si is over 500 ns at 60 K and decay in Yafet's $T^{-5/2}$ power law as a function of temperature, which is predicted for an indirect-gap semiconductor. Although spin relaxation mechanisms of the highly doped silicon, especially in relation to temperature dependence, are not elucidated as of yet, it is quite reasonable to suspect that both the interface- and the bulk-originated factors will influence the spin transport. This is the one of unsolved questions in Si spintronics. In highly doped Si, many groups have also reported the electrical spin injection and detection into Si by using the non-local (NL) magnetoresistance (MR) method.[6,7,8,9] In spite of much effort, the detail of spin transport characterizations in Si, such as the spin diffusion length, have not been reported. We have recently demonstrated electrical injection and detection into highly electron doped Si using MgO tunnel barrier, where the carrier concentration was about $1 \times 10^{20}$ cm$^{-3}$.[7] In this paper, we distinguish the temperature dependence of NL signals into the interface and bulk contributions through the Hanle-type precession measurement with the sufficiently reliable data.

The SOI substrate was annealed at 900 degrees C for 60 minutes to activate the dopant and to provoke re-crystallization after the phosphors (P) was implanted into Si. The electron concentration of $5 \times 10^{19}$ cm$^{-3}$ was determined by the Hall effect measurements.

The temperature dependence of resistivity exhibited metallic behavior. The Si surface was washed by using a dilute HF solution to remove the thin natural silicon oxide layer, and rinsed in a de-ionized water and isopropyl alcohol. After that, the Fe(13 nm)/MgO(0.8 nm) layers were formed on a silicon-on-insulator (SOI) substrate with (100)-plane by molecular beam epitaxy (MBE). The thickness of MgO(0.8 nm) was a minimally obtained as the homogeneous layer, which was determined by the transmission electron microscopy (TEM). We confirmed that $SiO_x$ and/or unexpected impurities did not exist at MgO/Si interface by TEM and energy dispersive X-ray spectroscopy (EDS). The resistance area (RA) product and channel resistivity at 100 μA and 8 K are estimated to be about 6 kΩμm$^2$ and $2 \times 10^{-3}$ Ωcm, respectively.

As shown in Fig.1, we fabricated the four-probe lateral spin devices by using electron beam lithography and a mesa-etching technique. First, Contact 2 ($0.5 \times 21$ μm$^2$) and Contact 3 ($2.5 \times 21$ μm$^2$) were defined by Ar ion milling. The edge to edge distance ($d_s$) between contact 2 and 3 are from 0.3 to 1.6 μm. Then, 30 nm-thick $SiO_2$ was deposited on the devices surface without the upper face of the electrodes. After that, the contact holes were formed into the $SiO_x$ and were filled in Al, which were contact 1 and 4. Finally, the pad electrodes on contact 1, 2, 3, and 4, Au (150 nm)/Cr (50 nm), were fabricated by lift-off methods.[7]

The NL MR measurements were performed using a standard AC lock-in technique ($f =$ 333 Hz) with an in-plane magnetic field along $y$-direction. The AC current was determined by using a DC voltage that was required for measuring DC currents for the contact between 1 and 2. Although the Hanle-type spin precession measurements is the same condition as the NL MR measurements, a magnetic field was applied along $z$-direction.

Fig. 2 shows the gap dependence of the NL output voltage ($\Delta V$) below 60 K with 500 μA. It is shown that the $\Delta V$ exponentially decays with the increasing of gap length, and the spin diffusion length ($\lambda_N$) determined by this gap dependence of the spin signals[7,10] is shown in the inset of Fig. 2; $\lambda_N$ is 2.35±0.53 μm at 20 K. Surprisingly, the $\lambda_N$ is almost constant up to 60 K in spite of the decrease of the output voltage with increasing temperature.

In order to understand the temperature effect more clearly, the Hanle-type spin precession of the NL MR measurements have been carried out, and the results are shown in Fig. 3 (a) 25 K, (b) 75 K and (c) 100 K. The signals could be clearly observed including the crossing point. The Hanle-type spin precession is expressed as a function of applied magnetic field along $z$-direction.[11]

$$\frac{\Delta V(B_\perp)}{I} = \pm \frac{P^2}{e^2 N(E_F) A} \int_0^\infty \varphi(t) \cos(\omega t) \exp(\frac{-t}{\tau_{sp}}) dt \ ,$$
$$\varphi(t) = \sqrt{1/4\pi Dt} \ \exp(-d^2/4Dt) \ ,$$

where $P$ is spin polarization, $D$ is the spin diffusion constant, and $\tau_{sp}$ is the spin life time, $\omega = g\mu_B B$ is the Larmor frequency, and $g$ is the $g$ factor of electrons (using $g = 2$). The data are in good agreement with the fitting function in every detail. We adopted the center to center distance of contact 2 and 3 as the gap length ($d_L$ = 2.33 μm) as shown in Fig. 1, because the electrode size is not explicitly included in the fitting procedure of the Hanle-type spin precession. In case of using $d_s$ = 0.83 μm, the $\lambda_N$ was underrepresented.[12] The parameters determined by the best fittings are as follows; $D$ = 6.50±0.20 cm$^2$/s, $\tau_{sp}$ = 9.21±0.23 ns, and $P$ = 0.016 at 25 K, respectively. We obtain the $\lambda_N$ = 2.45±0.07 μm  by using $\lambda_N = (D\tau_{sp})^{1/2}$, which is thus in good agreement with that by the gap dependence of the NL signals. Among the above parameters, we can take $\lambda_N$ and $\tau_{sp}$ as representing the character of the Si channel, and $P$ as representing the character of the interface between MgO and Si at the injector and detector.

Fig. 4 shows the temperature dependence of (a) $\Delta V$, (b) $P$, (c) $\lambda_N$, and (d) $\tau_{sp}$ by the Hanle measurement of up to 125 K. The $\Delta V$ exponentially decreases as a function of temperature. However, $\lambda_N$ decreases very gradually for the temperature evolution. Therefore, our results here indicate that this major origin of temperature dependence at $\Delta V$ does not exist in the Si channel. Concerning to the interface, $P$ lineally decreases with increasing temperature. Since the $\Delta V$ is proportional to the square of the spin polarization $P^2$, this steep reduction of $P$ results in a major factor of the temperature dependence of $\Delta V$.

The $\tau_{sp}$ estimated up to 125 K are shown in Fig. 4(d). According to the Elliott-Yafet theory[13,14], spin flip is caused by the impurity scattering such as the dopant as well as phonon scattering. In our case, the $\tau_{sp}$ is estimated to be 7.7 ns at 75 K, which is much smaller than that of the non-doped Si (~400 ns at the same temperature).[4] Compared to the Yafet's $T^{-5/2}$ power law of non-doped Si, our data shows a much slower decrease, which may originate from the metallic electronic transport in our samples where $R$(8 K)/$R$(300 K) was about 1/2. The $n$-type Si, with P implanted, is known to show the metal-insulator transition at the doping level of $3.74 \times 10^{18}$ cm$^{-3}$, and it behaves as a degenerate semiconductor at the doping level of more than $2 \times 10^{19}$ cm$^{-3}$.[14, 15] In our samples, since the carrier concentration is $5 \times 10^{19}$ cm$^{-3}$, Fermi surface is expected to exist at the conduction band. In a degenerate semiconductor, a part of factor $k_B T$ included in the scattering cross section can be replaced by the temperature independent Fermi energy.[13] So, the temperature dependence of $\tau_{sp}$ should be more gradual than Yafet's power law in a metallic sample. In this analogy, pure metals are expected to have a gradual temperature dependence of $\tau_{sp}$, and in fact, $\tau_{sp}$ of Cu was found to increase no more than 4 times between 300 and 50 K.[17]

In conclusion, we have reported the temperature dependence of the spin diffusion length in the highly doped Si channel ($5 \times 10^{19}$ cm$^{-3}$) with a lateral spin valve structure. The spin diffusion length estimated by the Hanle-type spin precession is in good agreement with that of the gap dependence of NL signals. By evaluating the interface and Si bulk channel separately, we demonstrate that the major factor of the temperature dependence at NL signals originates from reducing the spin polarization at the interface between the tunnel barrier and Si. It is possible that the interface state, such as the dangling-bond level, cause the spin scattering. The temperature dependence of $\tau_{sp}$ at the highly doped Si is smaller than that of the non-doped Si. This indicates that the highly doped Si has the different spin relaxation process from non-doped Si.

Acknowledgement

The authors are grateful to T. Takano, K. Muramoto and N. Mitoma (Osaka University) for valuable discussions and the experimental support and to M. Kubota, Y. Ishida, S. Tsuchida, and K. Yanagiuchi (TDK) for experimental support in sample analysis and fruitful discussions.

Figure captions

Fig.1 A cross-section diagram of the lateral device geometry. A magnetic field was applied along the long axis of the magnetic contact ($y$-direction) for NL MR measurements. With increasing the field, the magnetization of contact 2 and 3 are parallel or anti-parallel orientation. The Hanle-type spin presession is expressed as a function of applied magnetic field along $z$-direction.

Fig.2 The gap dependence of the output voltage and fitting results at any temperature. The data is normalized at 500 μA. The inset shows the temperature dependence of spin diffusion length.

Fig.3. Result of a Hanle-type spin precessions at (a) 25 K, (b) 75 K and (c) 100 K with $d_L = 2.33$ mm. The injection current was set at 1 mA. The red and blue open circles are experimental data, and the red and blue lines are the result of the fits. The arrows indicate the relative magnetization configuration (parallel or antiparallel) at contact 2 and 3.

Fig.4 The τεμπερατυρε δεπενδενχε οφ (α) τηε ουτπυτ ϖολταγε $\Delta V$, (b) the spin polarization $P$, (c) the spin diffusion length $\lambda_N$, (d) the spin lifetime $\tau_{sp}$, with the $d_L = 2.33$ μm up to 125 K by using Hanle-type spin precession.

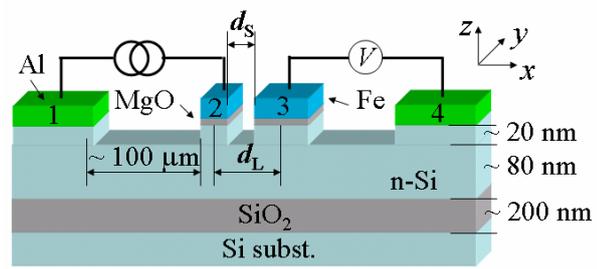

Figure. 1    T. Sasaki *et al.*

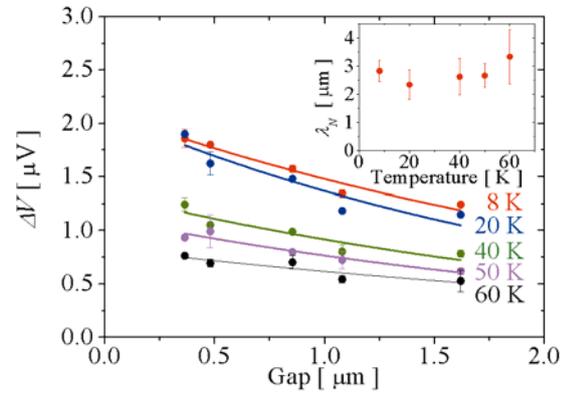

Figure. 2   T. Sasaki *et al*.

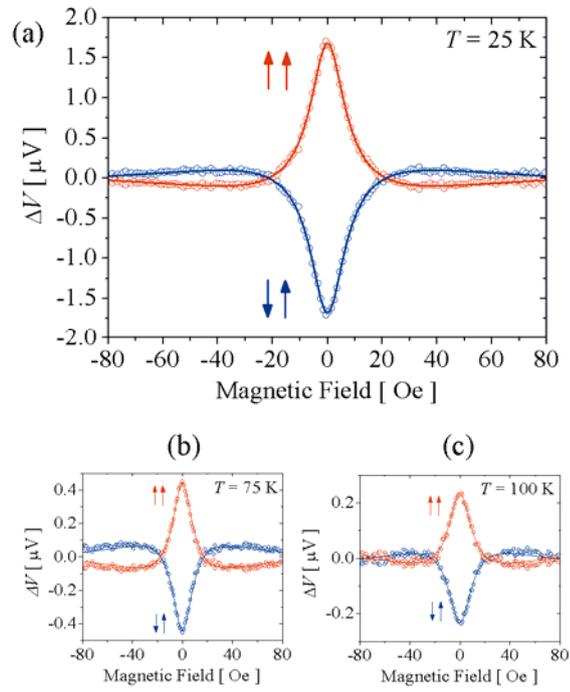

Figure. 3   T. Sasaki *et al*.

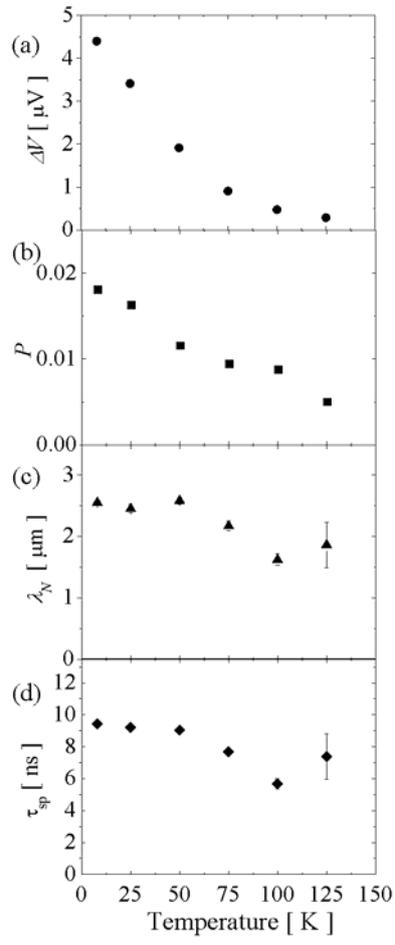

Figure. 4    T. Sasaki *et al*.